# On lamps, walls, and eyes: the spectral radiance field and the evaluation of light pollution indoors

**Salvador Bará**[1,*] and **Jaume Escofet**[2]

[1]*Área de Óptica, Departamento de Física Aplicada, Universidade de Santiago de Compostela, Santiago de Compostela, Galicia, Spain.*
[2]*Departament d'Òptica i Optometria, Universitat Politècnica de Catalunya, Terrassa, Catalunya, Spain.*
[*]*Corresponding author: salva.bara@usc.es*

**Abstract**

Light plays a key role in the regulation of different physiological processes, through several visual and non-visual retinal phototransduction channels whose basic features are being unveiled by recent research. The growing body of evidence on the significance of these effects has sparked a renewed interest in the determination of the light field at the entrance pupil of the eye in indoor spaces. Since photic interactions are strongly wavelength-dependent, a significant effort is being devoted to assess the relative merits of the spectra of the different types of light sources available for use at home and in the workplace. The spectral content of the light reaching the observer eyes in indoor spaces, however, does not depend exclusively on the sources: it is partially modulated by the spectral reflectance of the walls and surrounding surfaces, through the multiple reflections of the light beams along all possible paths from the source to the observer. This modulation can modify significantly the non-visual photic inputs that would be produced by the lamps alone, and opens the way for controlling -to a certain extent- the subject's exposure to different regions of the optical spectrum. In this work we evaluate the expected magnitude of this effect and we show that, for factorizable sources, the spectral modulation can be conveniently described in terms of a set of effective filter-like functions that provide useful insights for lighting design and light pollution assessment. The radiance field also provides a suitable bridge between indoor and outdoor light pollution studies.

# 1. Introduction

A growing body of research has underscored the role of light as a regulator of significant physiological processes, most notably the entrainment of the human circadian system [1-8]. The discovery in 2002 of the intrinsically photosensitive retinal ganglion cells (ipRGC), and of their anatomical and functional connections with the brain centers responsible for circadian regulation [9-15], provided a sound mechanistic basis for understanding several of the photic responses previously observed in laboratory settings and in clinical practice. This has led to a renewed interest in the quantitative determination of the light field at the entrance of the eye, particularly in indoor settings.

Humans in modern societies spend a considerable amount of time inside buildings, subjected to illumination levels and spectral distributions noticeably different from the ones existing in the natural environment. The progressive extension of the human activity into nighttime, enabled by the availability of artificial light sources and the low energy prices, has been identified as a likely cause of disruption of the circadian patterns in wide sectors of the population, with potentially significant health consequences [7,8]. Not surprisingly, there is a growing consensus about the need of evaluating the non visual effects of light when selecting light sources for indoor applications. Several physical magnitudes are of interest for assessing these effects: the absorbed dose, spectral composition, directionality, timing, previous photic history, and duration of the exposure are some of them. Particular attention has been given to the spectral content, since both the ipRGC excitation and the associated non-visual responses (measured, e.g., by the control-adjusted percentage of acute melatonin suppression after exposure to a light pulse at the central hours of the night) show a strong wavelength dependence, peaking at the short wavelength region of the visible spectrum [11,12].

The most conspicuous recent changes in indoor lighting are precisely those related to the spectral composition of light. Whilst other lighting parameters tend to evolve relatively slowly in time (for instance, the recommended illumination levels, or the typical timespan of the urban nighttime activities), the spectrum of the light sources is undergoing nowadays a fast and deep transformation. The widespread introduction of solid-state lighting (SSL) devices, particularly those based on phosphor-converted light emitting diodes (pc-LED), is altering the spectral landscape to which we were used after decades of lighting based on gas-discharge, fluorescent, and thermal sources. This fact explains the intense research effort devoted to assess the relative merits of these new sources in terms of their potential side-effects on the environment and human health [16-24].

Current phototransduction models [25-29] quantify the non-visual photic inputs in terms of appropriately weighted integrals of the spectral irradiance at the eye cornea, or, to be more precise, on a plane tangent to the cornea and perpendicular to the line of sight. Although this approach involves a certain simplification of the problem, because it does not take into account the spatial distribution of the radiance entering the eye [30], it was instrumental for enabling the first proposals of physiologically-based magnitudes describing non-visual inputs [25-29], and quantitative models for predicting the expected outputs after exposure to a light pulse under certain well defined experimental conditions [27-29]. Most of the above quoted research was directly applied to evaluate the photic effects of different types of lamp spectra. In typical indoor settings, however, the spectral composition of the light that actually reaches the observer's eyes does not depend on the characteristics of the light sources alone, but also on the spectral reflectance of the surrounding environment. Since this spectral reflectance is generally space-variant, the corneal irradiance will additionally depend on the direction of gaze. Any meaningful description of the actual light exposure conditions in indoor spaces must take into acoount the contribution of the surrounding surfaces, through the multiple reflections of the light beams in their way from the source to the observer.

In this paper we show that the spectral radiance field is a useful tool for a comprehensive description of the light entering the eye. In its seven-dimensional version it provides spatial, angular, spectral and time-resolved radiance information within any region of interest. Other radiometric and photometric magnitudes can be computed by integrating this field over the appropriate (solid angle, surface, wavelength and/or time) domains. We describe how the radiance field can be applied to the study of light pollution in indoor spaces, and propose a set of effective filter-like funtions useful for indoor lighting design and light pollution assessment.

## 2. The spectral radiance field in indoor spaces

The basic physical entity for describing radiative transfer processes at optical frequencies is the *spectral radiance field*, also called *plenoptic function* [31], denoted by

$$L = L(x, y, z; \theta, \varphi; \lambda; t) = L_\lambda(\mathbf{x}, \boldsymbol{\omega}, t) \qquad (1)$$

where $L_\lambda(\mathbf{x}, \boldsymbol{\omega}, t)$ is the spectral radiance [$Wm^{-2}sr^{-1}nm^{-1}$] at point $\mathbf{x} = (x, y, z)$ and time $t$, in the direction of the unit vector $\boldsymbol{\omega}(\theta, \varphi)$, at wavelength $\lambda$.

$L_\lambda(\mathbf{x}, \boldsymbol{\omega}, t)$ is a scalar field defined on a seven-dimensional space. However, this space can be reduced to five dimensions in rooms of ordinary size. A very short time after switching on the lights (~ tens of ns) the radiance distribution reaches a stationary state, and the time dependence can be safely dropped from Eq. (1). On the other hand, since the radiance along geometrical rays is conserved [32] –the attenuation by air molecules and aerosols within the room being negligible for all practical purposes– an additional dimension can be removed. The invariance of radiance implies that $L_\lambda(\mathbf{x}_O, \boldsymbol{\omega}) = L_\lambda(\mathbf{x}, \boldsymbol{\omega})$, that is, the radiance at any point $\mathbf{x}_o$ within the room, in the direction $\boldsymbol{\omega}$, is equal to the radiance of the wall (or source) point $\mathbf{x}$ located at the origin of the corresponding geometric ray (Fig. 1). In consequence, the radiance within the room is univocally determined by the radiance of its surfaces (walls, floor, ceiling, and the surfaces of any surrounding objects, including lamps).

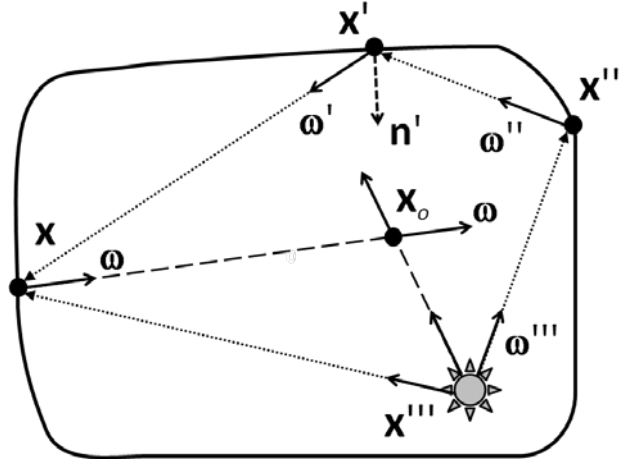

**Fig. 1.** The radiance field in indoor spaces. Notation: $\mathbf{x}_O$: arbitrary observation point; $\mathbf{x}, \mathbf{x'}, \mathbf{x''}, \mathbf{x'''}$: points on the walls and surfaces surrounding the observer, including the surface of the sources; $\boldsymbol{\omega}, \boldsymbol{\omega}', \boldsymbol{\omega}'', \boldsymbol{\omega}'''$: unit vectors in direction of the corresponding rays; $\mathbf{n'}$: unit vector normal to the surface at $\mathbf{x'}$ (the unit vectors $\mathbf{n}, \mathbf{n''}, \mathbf{n'''}$, normal to the surfaces at other points, are not drawn for clarity). The radiance at $\mathbf{x}_O$ in the direction $\boldsymbol{\omega}$ is equal to the radiance leaving $\mathbf{x}$ towards $\mathbf{x}_O$, $L_\lambda(\mathbf{x}_O, \boldsymbol{\omega}) = L_\lambda(\mathbf{x}, \boldsymbol{\omega})$. These two points are located along the same geometrical ray, and consequently fulfill the equation $\mathbf{x} = \mathbf{x}_O - \mu\boldsymbol{\omega}$ for some value of $\mu > 0 \in \mathfrak{R}$.

The radiance leaving $\mathbf{x}$ in the direction $\boldsymbol{\omega}$, $L_\lambda(\mathbf{x}, \boldsymbol{\omega})$, is, in turn, the sum of two terms: the radiance actively emitted by this point, $\tilde{L}_\lambda(\mathbf{x}, \boldsymbol{\omega})$, in case it belongs to a light source, and the fraction of the radiance $L_\lambda(\mathbf{x}, \boldsymbol{\omega}')$ that, coming from the remaining points of the room, is reflected at $\mathbf{x}$ (applicable to both, source surfaces and wall points). This is expressed by the well-known rendering equation [33]

$$L_\lambda(\mathbf{x}, \boldsymbol{\omega}) = \tilde{L}_\lambda(\mathbf{x}, \boldsymbol{\omega}) + \int_\Omega L_\lambda(\mathbf{x}, \boldsymbol{\omega}') f_\lambda(\mathbf{x}, \boldsymbol{\omega}', \boldsymbol{\omega}) \cos\alpha \, d\omega' \qquad (2)$$

where $f_\lambda(\mathbf{x}, \boldsymbol{\omega}', \boldsymbol{\omega})$ is the spatially-variant bidirectional reflectance distribution function (BRDF), ([32], p. 182), with units $sr^{-1}$, defined as the ratio of the radiance reflected at $\mathbf{x}$ in the direction $\boldsymbol{\omega}$ to the irradiance incident from an elementary solid angle $d\omega'$ around the unit vector $\boldsymbol{\omega}' = (\mathbf{x} - \mathbf{x'})/\|\mathbf{x} - \mathbf{x'}\|$:

$$f_\lambda(\mathbf{x}, \boldsymbol{\omega}', \boldsymbol{\omega}) = dL_\lambda(\mathbf{x}, \boldsymbol{\omega})/dE_\lambda(\mathbf{x}, \boldsymbol{\omega}') \qquad (3)$$

The integral in Eq.(2) is extended to $\Omega$, the whole hemisphere facing $\mathbf{x}$ whose pole lies in the direction of $\mathbf{n}$, the normal to the surface at that point. $\alpha$ is the angle of incidence measured from $\mathbf{n}$, that is, $\cos\alpha = (-\boldsymbol{\omega}')\cdot\mathbf{n}$. In the reference frame centered at $\mathbf{x}$ with the Z axis pointing in the direction of $\mathbf{n}$, the elementary solid angle in spherical coordinates is given by $d\omega' = \sin\theta\, d\theta\, d\phi$.

The classical solution of Eq. (2) for any given light source distribution can be straightforwardly obtained in terms of linear operators acting on the Hilbert space of the radiance field vectors, with a conventional integral inner product. To that end, let us denote by $\mathbf{L}$ and $\mathbf{L}_S$ the (infinite-dimensional) vectors composed of the values of $L_\lambda(\mathbf{x},\boldsymbol{\omega})$ and $\tilde{L}_\lambda(\mathbf{x},\boldsymbol{\omega})$, respectively, for all possible $(\mathbf{x},\boldsymbol{\omega})$. According to Eq. (2), both vectors are related by the equation

$$\mathbf{L} = \mathbf{L}_S + \mathbf{H}\mathbf{L}, \tag{4}$$

where $\mathbf{H}$ is the infinite-size matrix associated with the integral linear operator in Eq. (2). Regrouping terms and solving for $\mathbf{L}$ we get:

$$\mathbf{L} = (\mathbf{I} - \mathbf{H})^{-1}\mathbf{L}_S, \tag{5}$$

where $\mathbf{I}$ is the identity matrix. Finally, expanding the inverse as a Neumann operator series [34] we get the desired solution

$$\mathbf{L} = \sum_{n=0}^{\infty} \mathbf{H}^n \mathbf{L}_S. \tag{6}$$

The *n*-th term of this series corresponds to the contribution of the light that was emitted by the sources and underwent *n* diffuse reflections at the room surfaces. The existence of the inverse in Eq. (5) and the convergence of the series in Eq. (6) are guaranteed because the operator norm is smaller than one. To see it, let's define the norm in the $\mathbf{L}$ space as the total energy contained within the room, i.e.

$$\|\mathbf{L}\| = \frac{1}{c}\int_\lambda \int_{\mathbb{R}^3} \int_\Omega L_\lambda(\mathbf{x},\boldsymbol{\omega})\, d\boldsymbol{\omega}\, d^3\mathbf{x}\, d\lambda, \tag{7}$$

where $c$ is the speed of light. The condition $\|\mathbf{HL}\| \leq \gamma \|\mathbf{L}\|$, for some $\gamma < 1$, is automatically satisfied because the reflected radiance is always smaller than the incident one, due to unavoidable transmission and absorption losses at the surfaces.

An alternative deduction of Eq. (6) can be obtained by recalling that Eq. (4) is essentially a fixed-point equation to which the Banach contraction principle, also known as the Banach fixed-point theorem [35], can be applied. According to this theorem, the solution of Eq. (4) can be found as the limit of the series built by iteratively substituting it for $\mathbf{L}$ in its right-hand side, starting from any initial value (e.g. $\mathbf{L} = 0$). This leads directly to Eq. (6).

The explicit solution of Eq. (2) as a series of direct radiance from the sources and multiply scattered rays at the room surfaces can be easily deduced, although the notation can be cumbersome for all excepting the first terms. Note that Eq. (2) can be rewritten as

$$L_\lambda(\mathbf{x},\boldsymbol{\omega}) = \tilde{L}_\lambda(\mathbf{x},\boldsymbol{\omega}) + \int_\Omega L_\lambda(\mathbf{x}',\boldsymbol{\omega}') f_\lambda(\mathbf{x},\boldsymbol{\omega}',\boldsymbol{\omega}) \cos\alpha \, d\omega' \qquad (8)$$

because, due to the invariance of radiance, $L_\lambda(\mathbf{x},\boldsymbol{\omega}') = L_\lambda(\mathbf{x}',\boldsymbol{\omega}')$, i.e. the radiance incident on $\mathbf{x}$ in the direction $\boldsymbol{\omega}'$ is equal to the radiance leaving the point $\mathbf{x}'$ located at the origin of the ray. Now, an equation similar to Eq. (8) can be written for $\mathbf{x}'$:

$$L_\lambda(\mathbf{x}',\boldsymbol{\omega}') = \tilde{L}_\lambda(\mathbf{x}',\boldsymbol{\omega}') + \int_\Omega L_\lambda(\mathbf{x}'',\boldsymbol{\omega}'') f_\lambda(\mathbf{x}',\boldsymbol{\omega}'',\boldsymbol{\omega}') \cos\alpha' \, d\omega'' \qquad (9)$$

where $\cos\alpha' = (-\boldsymbol{\omega}'') \cdot \mathbf{n}'$. Substituting it for $L_\lambda(\mathbf{x}',\boldsymbol{\omega}')$ in Eq. (8), and applying repeatedly this procedure, leads to the series:

$$L_\lambda(\mathbf{x},\boldsymbol{\omega}) = \tilde{L}_\lambda(\mathbf{x},\boldsymbol{\omega}) + \int_\Omega \tilde{L}_\lambda(\mathbf{x}',\boldsymbol{\omega}') f_\lambda(\mathbf{x},\boldsymbol{\omega}',\boldsymbol{\omega}) \cos\alpha \, d\omega' +$$

$$\int_\Omega \left[ \int_\Omega \tilde{L}_\lambda(\mathbf{x}'',\boldsymbol{\omega}'') f_\lambda(\mathbf{x}',\boldsymbol{\omega}'',\boldsymbol{\omega}') \cos\alpha' \, d\omega'' \right] f_\lambda(\mathbf{x},\boldsymbol{\omega}',\boldsymbol{\omega}) \cos\alpha \, d\omega' +$$

$$\int_\Omega \left[ \int_\Omega \left[ \int_\Omega \tilde{L}_\lambda(\mathbf{x}''',\boldsymbol{\omega}''') f_\lambda(\mathbf{x}'',\boldsymbol{\omega}''',\boldsymbol{\omega}'') \cos\alpha'' \, d\omega''' \right] f_\lambda(\mathbf{x}',\boldsymbol{\omega}'',\boldsymbol{\omega}') \cos\alpha' \, d\omega'' \right] f_\lambda(\mathbf{x},\boldsymbol{\omega}',\boldsymbol{\omega}) \cos\alpha \, d\omega' + \ldots$$

$$(10)$$

The innermost integral of each term is effectively carried out over the solid angle $\Omega = \Omega_s$ subtended by the source from each point $\mathbf{x}, \mathbf{x'}, \mathbf{x''}$.... The remaining integrals are performed over the $\Omega = 2\pi$ hemispheres facing each of these points, since they correspond to the light diffusely reflected at the room surfaces. A more compact expression can be obtained by defining the linear operator series:

$$\begin{aligned}
G_0[\tilde{L}_\lambda(\mathbf{x'},\boldsymbol{\omega}');\mathbf{x},\boldsymbol{\omega}] &= \tilde{L}_\lambda(\mathbf{x},\boldsymbol{\omega}) \\
G_1[\tilde{L}_\lambda(\mathbf{x'},\boldsymbol{\omega}');\mathbf{x},\boldsymbol{\omega}] &= \int_\Omega \tilde{L}_\lambda(\mathbf{x'},\boldsymbol{\omega}') f_\lambda(\mathbf{x},\boldsymbol{\omega}',\boldsymbol{\omega}) \cos\alpha \, d\omega' \\
&\vdots \\
G_n[\tilde{L}_\lambda(\mathbf{x'},\boldsymbol{\omega}');\mathbf{x},\boldsymbol{\omega}] &= G_1[G_{n-1}[\tilde{L}_\lambda(\mathbf{x'},\boldsymbol{\omega}');\mathbf{x},\boldsymbol{\omega}]]
\end{aligned} \qquad (11)$$

and rewritting Eq. (10) as:

$$L_\lambda(\mathbf{x},\boldsymbol{\omega}) = \sum_{n=0}^{\infty} G_n[\tilde{L}_\lambda(\mathbf{x'},\boldsymbol{\omega}');\mathbf{x},\boldsymbol{\omega}]. \qquad (12)$$

$G_n[\tilde{L}_\lambda(\mathbf{x'},\boldsymbol{\omega}');\mathbf{x},\boldsymbol{\omega}]$ is the operator that assigns to the radiance emitted by the source(s) the radiance that leaves $\mathbf{x}$ in the direction $\boldsymbol{\omega}$ after having undergone exactly $n$ reflections at the room surfaces (including the last one, at $\mathbf{x}$ itself).

Note that the general expression of Eq. (8) may adopt particular forms, depending on whether the point $\mathbf{x}$ belongs to a passive surface or to the sources. For points on the walls and other passive reflecting surfaces we have $\tilde{L}_\lambda(\mathbf{x},\boldsymbol{\omega}) = 0$: the radiance leaving these points is just built up from the diffuse reflections at them of the light coming from the sources and the remaining wall points, as indicated by the integral term. For points belonging to the sources, in turn, $\tilde{L}_\lambda(\mathbf{x},\boldsymbol{\omega}) \neq 0$, and both terms of Eq. (8) have to be taken into account (the integral term accounts for the reflections at the source surface). When the light actively emitted by a source is significantly more intense that the light reflected by its surface, as it is the case in many practical situations, the first term becomes dominant and, to a good approximation, we can set $L_\lambda(\mathbf{x},\boldsymbol{\omega}) \approx \tilde{L}_\lambda(\mathbf{x},\boldsymbol{\omega})$ in Eq. (8) for the source points.

For several indoor applications it is necessary to compute the spectral irradiance $E_\lambda(\mathbf{x}_o,\mathbf{n}_o)$ at a generic point $\mathbf{x}_o$ at the entrance of the eye of an observer whose line of sight is oriented along the unit vector $\mathbf{n}_o$. This irradiance is given by the cosine-weighted solid-angle integral of the incident radiance. The incident radiance is, in turn, equal to the radiance at the initial point of the ray, $L_\lambda(\mathbf{x}_o,\boldsymbol{\omega}) = L_\lambda(\mathbf{x},\boldsymbol{\omega})$. Hence:

$$E_\lambda(\mathbf{x}_o,\mathbf{n}_o) = \int_\Omega L_\lambda(\mathbf{x}_o,\boldsymbol{\omega})\cos\alpha_o\, d\omega = \int_\Omega L_\lambda(\mathbf{x},\boldsymbol{\omega})\cos\alpha_o\, d\omega \qquad (13)$$

where $\cos\alpha_o = (-\boldsymbol{\omega})\cdot\mathbf{n}_o$. Substituting Eq. (12) into Eq. (13) we get:

$$E_\lambda(\mathbf{x}_o,\mathbf{n}_o) = \sum_{n=0}^{\infty}\left[\int_\Omega G_n[\tilde{L}_\lambda(\mathbf{x}',\boldsymbol{\omega}');\mathbf{x},\boldsymbol{\omega}]\cos\alpha_o\, d\omega\right] \equiv \sum_{n=0}^{\infty} F_n[\tilde{L}_\lambda(\mathbf{x}',\boldsymbol{\omega}');\mathbf{x}_o,\mathbf{n}_o] \qquad (14)$$

where we have defined:

$$F_n[\tilde{L}_\lambda(\mathbf{x}',\boldsymbol{\omega}');\mathbf{x}_o,\mathbf{n}_o] = \int_\Omega G_n[\tilde{L}_\lambda(\mathbf{x}',\boldsymbol{\omega}');\mathbf{x},\boldsymbol{\omega}]\cos\alpha_o\, d\omega \qquad (15)$$

The linear operator $F_n$ is the result of an additional cosine-weighted integration of the operator $G_n$ over the $\Omega = 2\pi$ hemisphere centered at $\mathbf{x}_o$ with pole in the direction $\mathbf{n}_o$. $F_n$ gives the contibution of the *n*-times reflected rays to the overall irradiance at the observation point.

## 3. Factorizable sources: effective filter functions

Let us now assume that the sources are factorizable, in the sense defined in [36], i.e. that their spectral signature is the same for all active emitting points and all emission directions. Most practical indoor sources are factorizable to a good degree of approximation. For factorizable sources we have:

$$\tilde{L}_\lambda(\mathbf{x},\boldsymbol{\omega}) = \Phi(\lambda) B(\mathbf{x},\boldsymbol{\omega}) \qquad (16)$$

Both factors in Eq. (16) can be arbitrarily scaled, as long as their product give the correct value for the emitted radiance. A useful choice is to identify $\Phi(\lambda)$ with the overall spectral flux [W nm$^{-1}$] emitted by the source, in which case $B(\mathbf{x},\boldsymbol{\omega})$ turns out to be a function with units [m$^{-2}$sr$^{-1}$]. Since the linear operators $F_n$ and $G_n$ act on the geometrical coordinates but not on the wavelength, we immediately get:

$$L_\lambda(\mathbf{x},\boldsymbol{\omega}) = \Phi(\lambda)\sum_{n=0}^{\infty} G_n[B(\mathbf{x'},\boldsymbol{\omega'});\mathbf{x},\boldsymbol{\omega}] = \Phi(\lambda)S(\lambda;\mathbf{x},\boldsymbol{\omega}) \qquad (17)$$

$$E_\lambda(\mathbf{x}_o,\mathbf{n}_o) = \Phi(\lambda)\sum_{n=0}^{\infty} F_n[B(\mathbf{x'},\boldsymbol{\omega'});\mathbf{x}_o,\mathbf{n}_o] = \Phi(\lambda)T(\lambda;\mathbf{x}_o,\mathbf{n}_o) \qquad (18)$$

where we have defined the *Effective inverse surface solid angle function* [m$^{-2}$sr$^{-1}$] as:

$$S(\lambda;\mathbf{x},\boldsymbol{\omega}) = \sum_{n=0}^{\infty} G_n[B(\mathbf{x'},\boldsymbol{\omega'});\mathbf{x},\boldsymbol{\omega}] \qquad (19)$$

and the *Effective inverse surface function* (*EISF*) [m$^{-2}$] as:

$$T(\lambda;\mathbf{x}_o,\mathbf{n}_o) = \sum_{n=0}^{\infty} F_n[B(\mathbf{x'},\boldsymbol{\omega'});\mathbf{x}_o,\mathbf{n}_o] \qquad (20)$$

Equations (17) and (18) show that both the radiance at an arbitrary point of the room and the irradiance on an arbitrarily oriented surface can be expressed as the result of two effective filter-like functions, $S(\lambda;\mathbf{x},\boldsymbol{\omega})$ and $T(\lambda;\mathbf{x}_o,\mathbf{n}_o)$, respectively, acting on the source spectrum $\Phi(\lambda)$. These functions depend on the position and orientation of the observing point within the room, $(\mathbf{x}_o,\mathbf{n}_o)$, and on the spatial and angular emitting features of the source, through the function $B(\mathbf{x},\boldsymbol{\omega})$, according to Eqs. (19) and (20), but do not depend on the source spectrum. The wavelength enters into their argument through the spectral reflectance of the walls and other reflecting surfaces surrounding the observer. $S(\lambda;\mathbf{x},\boldsymbol{\omega})$ and $T(\lambda;\mathbf{x}_o,\mathbf{n}_o)$ describe the modulation of the original source spectrum due to the multiple reflections of the light beams in their way from the source to the observer.

Once determined, $S(\lambda;\mathbf{x},\boldsymbol{\omega})$ and $T(\lambda;\mathbf{x}_o,\mathbf{n}_o)$ can be used to evaluate the radiance and irradiance produced by sources with the same $B(\mathbf{x},\boldsymbol{\omega})$ but different spectra. This allows for an easy evaluation of the effects that can be expected if the spectral composition of the sources is changed, everything else kept equal, and for choosing the best advisable sources for each particular application. Note finally that these functions are not *filters* in the strict sense of this term, since they relate different physical magnitudes and hence they are not dimensionless.

**4. Experimental results**

In order to determine experimentally the order of magnitude of the spectral modulation effects due to the surroundings, we conducted a series of measurements of the spectral irradiance at the entrance plane of the eye of an observer, using different lamps and different directions of gaze, in a meeting room located within the premises of the Terrassa School of Optics and Optometry of Universitat Politècnica de Catalunya.

*A. Room distribution and wall reflectances*

The room was a space 6.00 m long, 2.37 m wide, and 2.55 m high, fitted with a meeting table at 0.735 m above the floor, a few chairs and scarce additional furniture (Fig. 2). Three of the walls, labeled South, East and West, were predominantly of a light orange hue and the remaining one (North) had a bigger proportion of dark orange tones. The wall labels were chosen for convenience and do not necessarily represent the actual geographical orientations. The lamps were located at the center of the room, hanging from the ceiling at 1.53 m above the table, and the observer was located 0.8 m off-center towards the North wall. When looking towards the South wall, both the direct light of the source and the reflected light from the walls arrive to the observer's eye. Only reflected light arrives to the eye when looking northwards.

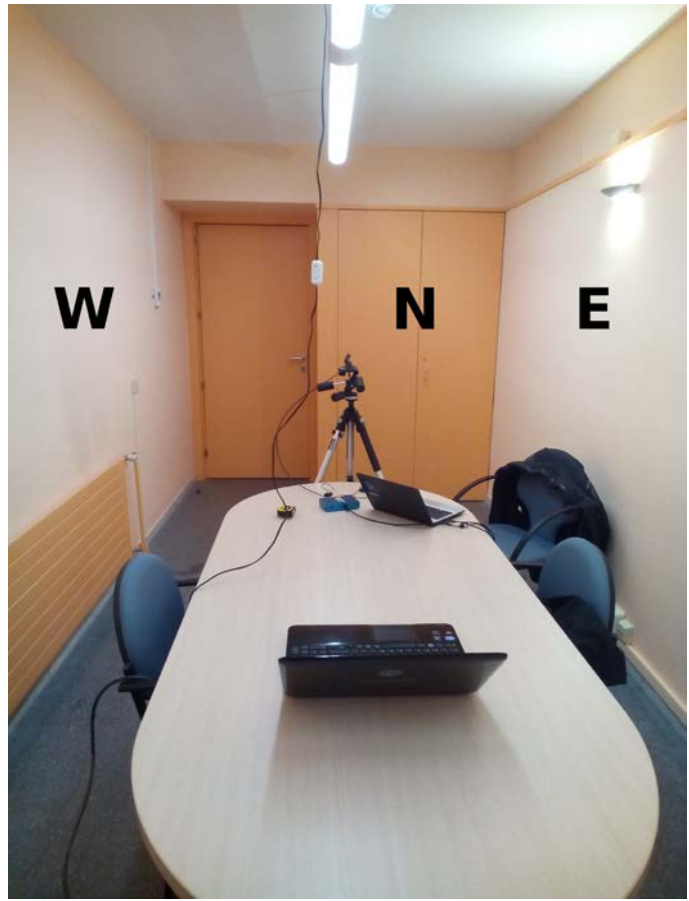

**Fig. 2.** Indoor space where the irradiance measurements were taken. The spectrometer head was located on the tripod shown in the figure, pointing horizontally. The walls visible in this figure are labeled W (West), N (North), and E (East).

The spectral reflectances of the walls, floor, ceiling, and table surface were measured with a SpectraScan PR-715 spectroradiometer (Photo Research Inc., Chatsworth, CA), and are shown in Figure 3. The dominant wall pigments act as long-pass filters in the wavelength domain, significantly attenuating, although with different strengths, the wavelenghts below 560 nm. The ceiling was painted matt white.

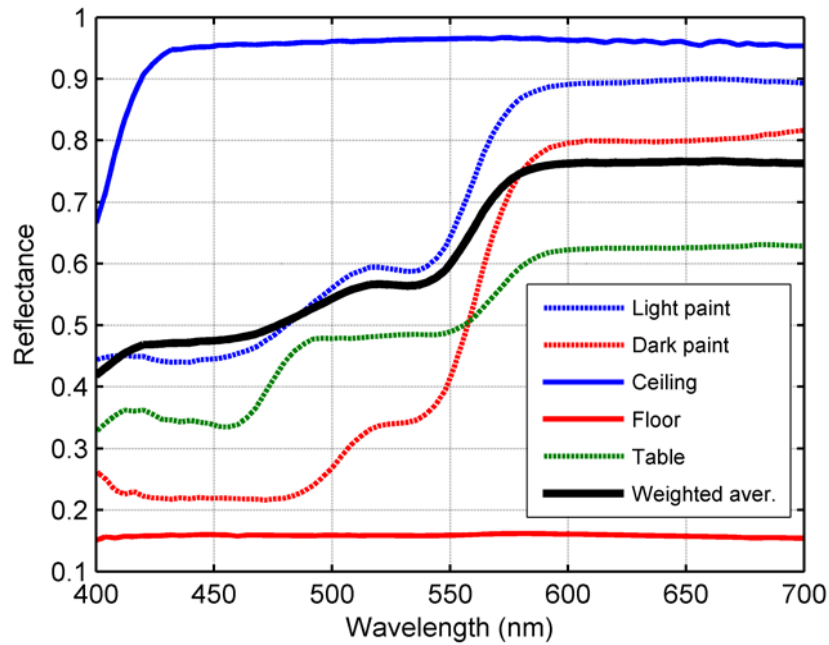

**Fig. 3.** Spectral reflectance of the room surfaces. Light paint: South, East and West walls, painted with light orange pigments; Dark Paint: North wall, darker orange paint. The weighted average of these spectral reflectances is used in Section 5 (see text for details).

*B. Lamp spectra*

Several solid-state (LED), halogen (HAL) and fluorescent (CFL) lamps were used for the measurements. The lamp set included three phosphor-converted white LEDs with correlated color temperatures (CCT) of 5000K (8 W, 600 lm), 4000K (8 W, 600 lm) and 2700K (10W, 600 lm), one halogen lamp of 2800K (46 W, 702 lm) and two CFL with CCT 6500K (24W, 1398 lm) and 2700K (13W, 664 lm). Their spectra were previously measured using a calibrated BlueWave spectrometer operating in the 350-1150 nm range (STE-BW-VIS-25, StellarNet, Inc, FL) with the entrance fiber located at 50 cm from the lamps, fitted with cosine correction. Its 25 μm slit provides a nominal spectral resolution of 1 nm. The spectral radiant flux of the lamps, normalized to a photopic flux of 1000 lm, is shown in Fig. 4.

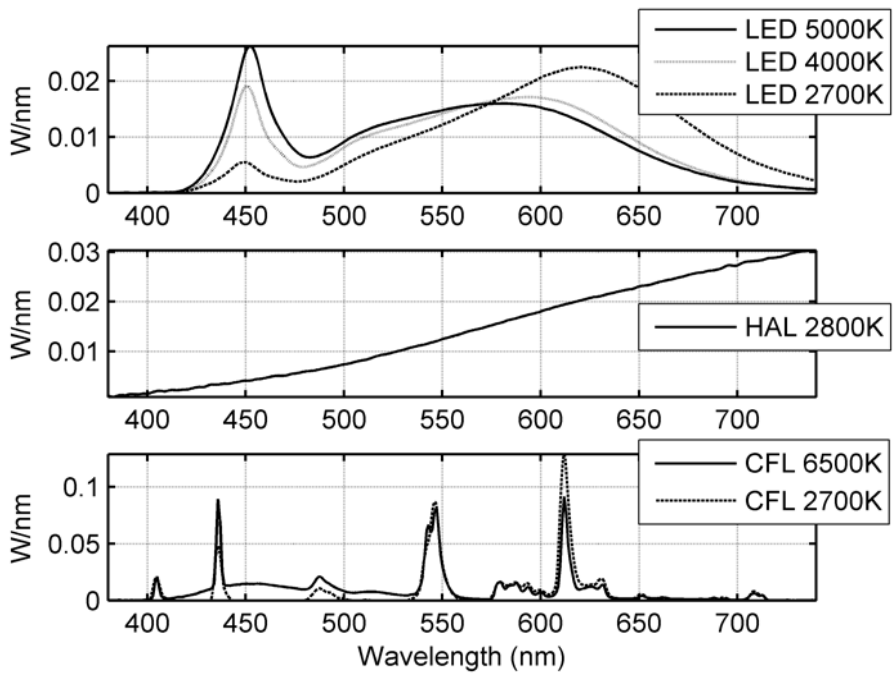

**Fig. 4.** Spectral radiant flux (W·nm$^{-1}$) of the lamps used in this study, normalized to a standard 1000 lm output.

*C. Measurement protocol*

The corneal spectral irradiances were measured in vertical planes at the position corresponding to the eye of an hypothetical observer sat at the end of the table, at 125 cm height, and pointing successively to the four walls of the room. The cosine-corrected fiber head of the STE-BW-VIS-25 spectrometer was attached to an altazimuth mount, and successive records of the spectral irradiance were taken with the device pointing perpendicularly towards each wall.

*D. Non-visual input/output magnitudes*

From the spectral irradiance recorded at each of the four gaze directions we computed the photoreceptoral inputs in the cyanopic, chloropic, erythropic, melanopic and rhodopic spectral bands (corresponding to the retinal S, L, and M cones, ipRGC, and rods, respectively), expressed in band-weighted W·m$^{-2}$ according to the CIE-normalized versions of the basic photoreceptor spectral sensitivities described in [25-26]. We also computed the circadian input ($CL_A$) and the expected melatonin suppression ratio (*CS*) using the model of Rea *et al* [27-29]. We additionally wanted to check an intriguing possibility, related to a central feature of this model: the possible existence of sign reversals of the blue-yellow spectral opponency term, due to the modulating effect of the room walls, relative to the sign of the direct light of the source.

*E. Results*

*1. Corneal irradiances*

The corneal irradiances were measured on a vertical plane at the center of the observer's pupil, for the six lamps and the four gaze directions included in this study. Figure 5 displays the results for the direction of gaze corresponding to the West wall. Comparing the shape of these spectral distributions with the ones emitted by the sources (Figure 4), the attenuating effect of the spectral reflectance of the walls on the short wavelength region of the visible spectrum can be clearly noticed.

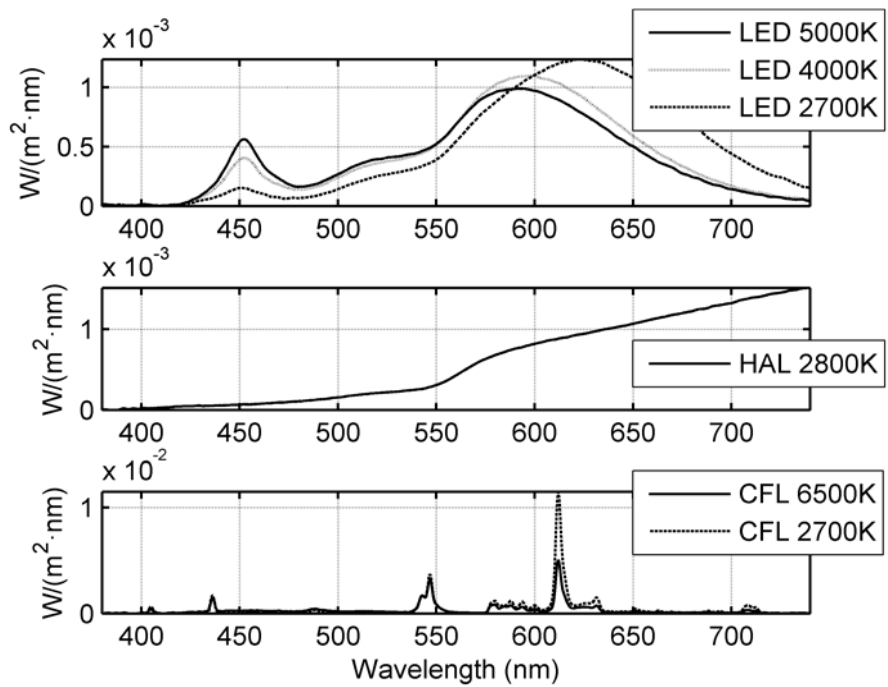

**Fig. 5.** Spectral irradiance, in $W \cdot m^{-2} \cdot nm^{-1}$, measured at the eye cornea of an observer looking at the West wall of the indoor space analyzed in this study. As in Fig. 4, the measured distributions were scaled for standard lamp outputs of 1000 lm.

*2. Effective inverse surface function*

The ratio of the corneal spectral irradiance to the spectral flux of the lamp is what we called in section 2 the effective inverse surface function, $T$, measured in units m$^{-2}$. Fig. 6 displays at the same scale the shape of the $T$ function for the LED and halogen lamps in the four gaze directions. The experimental functions corresponding to the CFL lamps were not included in this figure, since the wide spectral regions in which the flux emitted by the CFL is close to zero give rise to a strong noise propagation to the final $T$ profile. LED and halogen lamps, having relatively smooth and continuous spectra without zeros (Fig. 4), enable more stable experimental determinations of the $T$ function.

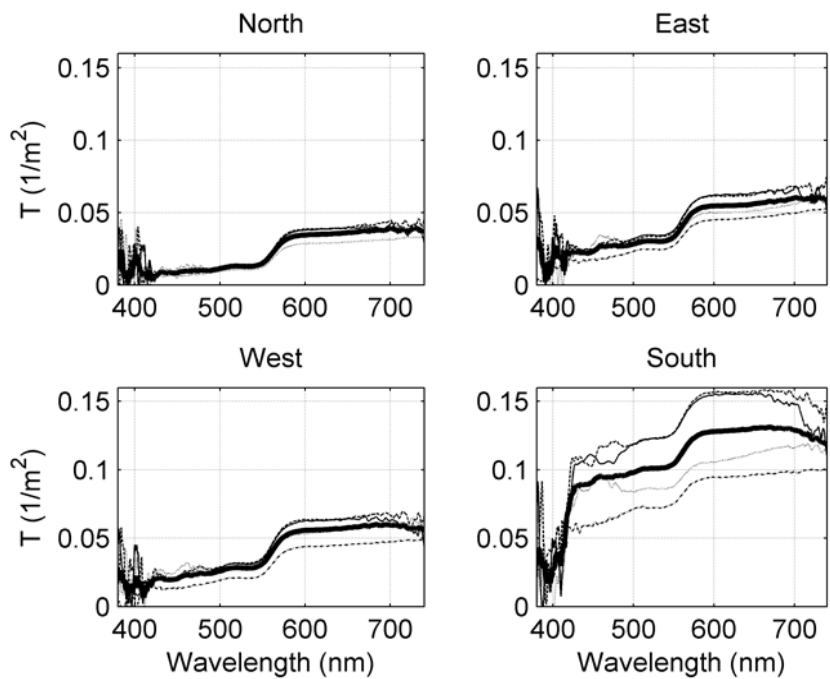

**Fig. 6.** The effective inverse surface functions $T$ (EISF), in units m$^{-2}$, for an observer in the conditions of the present experiment looking towards the four walls of the room. Sources: LED 5000K (thin full line), LED 4000K (dashed), LED 2700K (dotted), halogen 2800K (dash-dotted). The thick full line is the average of the individual plots.

From the plots displayed in Fig. 6 it can be seen that the shape of the function $T$ approaches the shape of the spectral reflectance of the walls for the gaze direction

North (when the source lies behind the observer and the light reaching the eye mainly stems from diffuse reflections at the North wall), and that the spectral contrast becomes less pronounced for the gaze direction South (when the source is in front of the observer, and a considerable amount of light arrives to the eye without undergoing reflections). The East and West directions represent intermediate situations, with some minor differences caused, among other factors, by the existence of a white surface in the E wall, and a dark orange region in the W.

*3. Photoreceptoral inputs*

The CIE-normalized inputs to the five basic photoreceptors of the human eye (S, L and M cones, intrinsically photosensitive retinal ganglion cells, and rods) were computed from the corneal irradiance by integration using the appropriate spectral weighting functions [25-26]. In order to evaluate the effects associated with the changing composition of the spectrum of the light arriving to the eye in different gaze directions, isolating them from the effects due to the change in the overall illuminance, these calculations were performed by scaling the irradiances at the entrance of the eye to a constant level of photopic illuminance (300 lx). Note that in an actual situation with a fixed lamp lumen output the photopic illuminance will in general be different for each viewing direction.

The results for the six lamps and the four gaze directions are graphically displayed in Figure 7. The scale bar units are $W \cdot m^{-2}$. The inputs corresponding to the direct observation of the source alone (i.e. in a space without walls and other reflecting surfaces), for the same level of corneal illuminance, are also displayed. In each matrix the rows, from top to bottom, correspond to the inputs due to the lamp alone and to the overall light received by the eye when gazing towards the South, East, West and North walls, respectively; the columns, from left to right, to the cyanopic, melanopic, rhodopic, chloropic and erythropic bands.

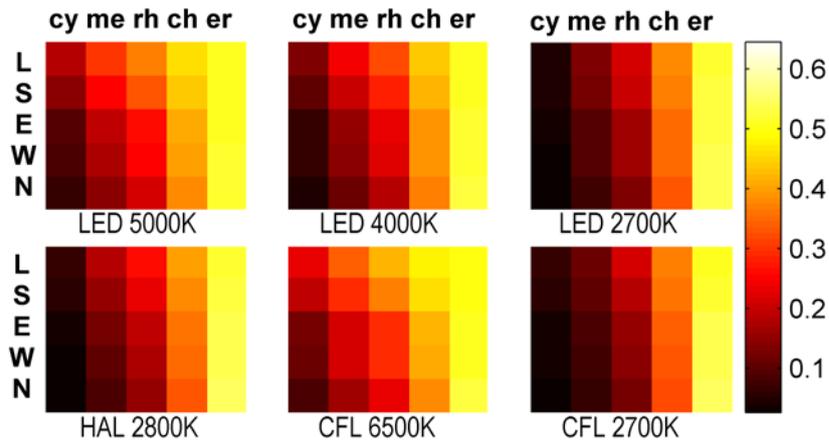

**Fig. 7.** CIE-normalized photoreceptoral inputs (W·m$^{-2}$) for several lamps and gaze directions, for a constant corneal illuminance of 300 lx. Top, left to right: LED lamps of 5000K, 4000K and 2700K. Bottom, left to right: Halogen of 2800K , CFL of 6500K, and CFL of 2700K. The rows of each data matrix correspond to the inputs due to the lamp alone (L), and to the overall light received by the eye gazing towards the South (S), East (E), West (W) and North (N) walls, respectively; the columns, to the cyanopic (cy), melanopic (me), rhodopic (rh), chloropic (ch) and erythropic (er) bands.

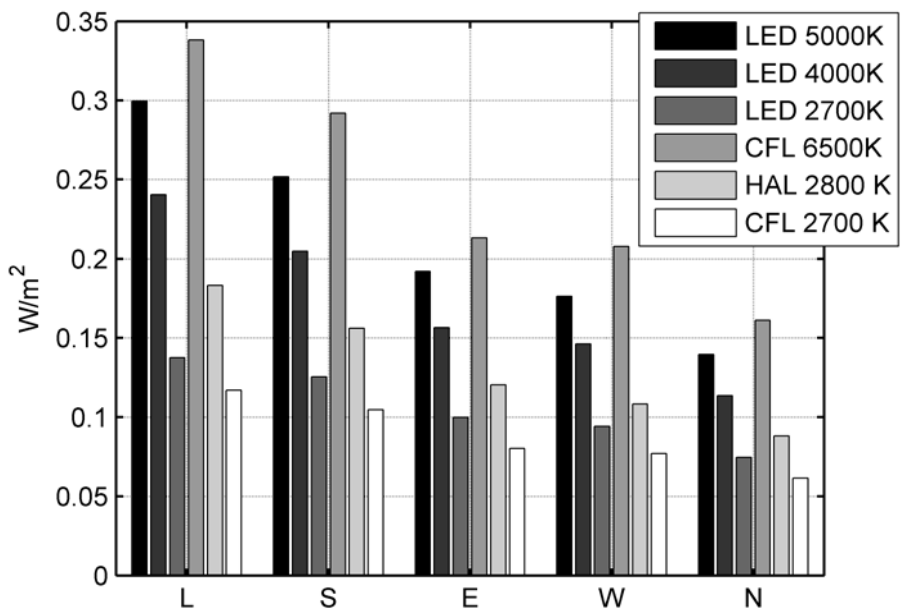

**Fig. 8.** CIE-normalized ipRGC photoreceptoral inputs (W·m$^{-2}$) for several lamps and gazing directions, at a constant corneal illuminance of 300 lx. The data groups correspond to the overall light received by the eye when gazing towards the lamp alone (L), and to the South (S), East (E), West (W) and North (N) walls.

Fig. 8 shows the inputs to the ipRGC (melanopic) channel, central to the circadian regulation system, for the different lamps and gaze directions. Both Fig. 7 and 8 show that the use of wall paints that attenuate the short wavelength region of the visible spectrum may be instrumental for reducing substantially the overall input to the ipRGCs. With our experimental settings this reduction amounts to a factor of about 0.5 for the N gaze direction, in comparison with the input that would be produced by the direct spectrum of the light source, at equal corneal illuminance.

*4. Circadian input ($CL_A$) and melatonin suppression ratio (CS)*

The results for the six sources and five gaze directions (staring directly at the source alone, and gazing to the four room walls) are summarized in Table 1. As done in the previous subsection, the corneal irradiances were normalized to a photopic illuminance of 300 lx, in order to decouple the changes due to the spectral composition of the light from the ones due to the differences in the average luminance of the walls, that give rise to different levels of overall illumination at the entrance of the eye.

**Table 1:** Circadian input ($CL_A$) and melatonin suppression ratio (CS, in %) [27-29] for the lamps and gazing directions considered in this work, under corneal photopic illuminance of 300 lx.

| | $CL_A$ (W/m$^2$) / CS (%) | | | | |
|---|---|---|---|---|---|
| Lamp | Direct | South | East | West | North |
| CFL 6500K | 400.2 / 37 | 309.7 / 32 | 148.6 / 19 | 140.1 / 18 | 251.7 / 28 |
| LED 5000K | 306.5 / 32 | 213.2 / 25 | 299.8 / 32 | 273.8 / 30 | 215.5 / 26 |
| LED 4000K | 179.9 / 22 | 318.4 / 33 | 240.6 / 28 | 224.4 / 26 | 172.9 / 22 |
| HAL 2800K | 277.6 / 30 | 235.6 / 27 | 180.1 / 22 | 161.6 / 21 | 131.3 / 17 |
| LED 2700K | 205.7 / 25 | 188.3 / 23 | 149.5 / 19 | 140.9 / 19 | 111.1 / 15 |
| CFL 2700K | 179.6 / 22 | 160.3 / 21 | 121.8 / 16 | 117.0 / 16 | 92.7 / 13 |

The behavior of the sign of the blue-yellow spectral opponency term, central to Rea et al. model, is shown in Table 2. With our experimental settings, the balance for the lower CCT sources (2700K-2800K) is always negative, irrespectively of the gaze direction. The light reaching the cornea from the remaining sources (CCT 6500K, 5000K, 4000K) undergoes a sign reversal when the spectral reflectance of the walls modifies its spectral composition in sufficient degree. Let us remind that in the South direction a considerable amount of light reaches the cornea directly from the source, and the rest comes mainly from reflections at walls with light orange reflectance; in the North direction, however, the lamp lies behind the observer head, and the only light reaching the cornea is the one diffusely reflected, especially at the relatively dark North wall (see Figure 2). Note that the sign of the spectral opponency term does not depend on the irradiance normalization, since it is determined by the shape of the spectral distribution and not by its absolute value.

**Table 2:** Sign of the blue-yellow spectral opponency term of the Rea *et al.* model [27-29]

| Lamp | Direct | South | East | West | North |
|---|---|---|---|---|---|
| CFL 6500K | + | + | + | + | – |
| LED 5000K | + | + | – | – | – |
| LED 4000K | + | – | – | – | – |
| HAL 2800K | – | – | – | – | – |
| LED 2700K | – | – | – | – | – |
| CFL 2700K | – | – | – | – | – |

*5. Correlated Color Temperature (CCT)*

Table 3 shows the CCT of the light arriving to the cornea of the eye under the observing conditions described above. Although CCT is a very imprecise metric for assessing circadian effects, it is widely used for reporting color appearance. From the results reported in this Table it can be seen that the wall reflectance can modify substantially the CCT of the light at the observer location and direction of gaze, in comparison with the one that would be adscribed to the direct light of the source.

**Table 3:** Correlated Color Temperature (K) of the light arriving to the observer eyes

| Lamp | Direct | South | East | West | North |
|---|---|---|---|---|---|
| CFL 6500K | 6162 | 5041 | 3286 | 3181 | 2437 |
| LED 5000K | 5056 | 4087 | 3164 | 2962 | 2521 |
| LED 4000K | 3977 | 3370 | 2721 | 2609 | 2251 |
| HAL 2800K | 2719 | 2378 | 2029 | 1930 | 1785 |
| LED 2700K | 2501 | 2264 | 1991 | 1941 | 1781 |
| CFL 2700K | 2658 | 2193 | 1848 | 1805 | 1662 |

## 5. A simple analytical model

Some insights on the role of the surrounding surfaces as modulators of the source spectrum, and, in particular, on the qualitative behaviour of the EISF shown in Fig. 6, can be obtained using a simplified model for which an analytical solution of Eq. (10) exists. Note that this highly simplified model is developed mainly as a complementary thinking aid, and it is not intended to be a substitute for the precise measurements or the numerical evaluations of the EISF function.

Let us consider a spherical enclosure of radius $R$, with homogeneous and Lambertian walls, and a small, almost pointlike, homogeneous, and isotropic light source located at its center (Fig. 9). Let us also consider an observer located at $\mathbf{x}_O$, a distance $R_O$ from the source, whose line of sight (perpendicular to the corneal plane) points in the direction $\mathbf{n}_O$. This set of assumptions implies that:

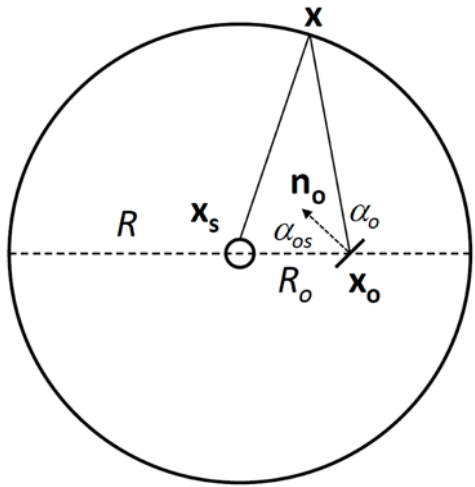

**Fig. 9.** An idealized spherical room of radius $R$ with a small spherical source at its center. The observer's eye is located at $\mathbf{x}_O$, a distance $R_O$ from the source, and its line of sight (perpendicular to the corneal plane) points in the direction $\mathbf{n}_O$. $\mathbf{x}$ is a generic point of the room. $\alpha_O$ is the polar angle of any arbitrary point of the walls or the source surface, as seen from the observer and measured with respecto to $\mathbf{n}_O$. $\alpha_{os}$ is the polar angle of the center of the source.

i) The polar angles of the points $\mathbf{x}_S$ of the source, measured at $\mathbf{x}_O$ with respect to the normal $\mathbf{n}_O$, are for all practical purposes equal to $\alpha_{os}$, the polar angle of the source center.

ii) All rays from the source incide perpendicularly on every point $\mathbf{x}$ of the enclosure. That is, for all $\mathbf{x}_S$ and $\mathbf{x}$, we have, to a good approximation, $\alpha \approx 0$ and $\cos\alpha \approx 1$, where $\alpha$ is the angle of incidence on $\mathbf{x}$, measured from its normal.

iii) The source radiance is constant for all points of the source and for all emission directions, and only depends on wavelength. That is, $\tilde{L}_\lambda(\mathbf{x}_S, \boldsymbol{\omega}_S) = \tilde{L}_\lambda$, and the source is trivially factorizable.

iv) For all points $\mathbf{x}$ of the spherical enclosure, and for all incident and diffusely reflected ray directions, $\boldsymbol{\omega}'$ and $\boldsymbol{\omega}$, the spectral bidirectional reflectance distribution function has the simple form $f_\lambda(\mathbf{x}, \boldsymbol{\omega}', \boldsymbol{\omega}) = \rho(\lambda)/\pi$ sr$^{-1}$, where $\rho(\lambda)$ is the spectral reflectance of the walls and the source limiting surfaces.

The radiance $\tilde{L}_\lambda$ emitted by the source can be easily determined from the source spectral flux, $\Phi_S(\lambda)$. This flux is equal (by definition) to:

$$\Phi_S(\lambda) = \int_S \int_\Omega \tilde{L}_\lambda(\mathbf{x}_S, \boldsymbol{\omega}_S) \cos\alpha_S \, dS \, d\omega_S, \qquad (21)$$

where the integral in $dS$ is extended to all points $\mathbf{x}_S$ of the source surface ($S$) and, for each $\mathbf{x}_S$, to all possible emission directions $\boldsymbol{\omega}_S$ contained in the half-sphere $\Omega$ of size $2\pi$ sr around the normal to the source surface at that point, being $\alpha_S$ the angle formed by these directions and the normal. Since the source radiance is assumed to be equal for all points and directions, and the elementary solid angle is given by $d\omega_S = \sin\alpha_S \, d\alpha_S \, d\phi$, we obtain the well-known formula for the spectral flux of Lambertian sources:

$$\Phi_S(\lambda) = \tilde{L}_\lambda \int_S dS \int_\Omega \cos\alpha_S \, d\omega_S = \tilde{L}_\lambda \, S \int_0^{2\pi} d\phi \int_0^{\pi/2} \cos\alpha_S \sin\alpha_S d\alpha_S = \pi \, \tilde{L}_\lambda \, S \qquad (22)$$

and hence:

$$\tilde{L}_\lambda = \frac{\Phi_S(\lambda)}{\pi S} \qquad (23)$$

Note that under the above quoted hypotheses, the integral $\int_{\Omega_S} \tilde{L}_\lambda(\mathbf{x}_S, \boldsymbol{\omega}) \cos \alpha_o \, d\omega$, computed from the observer position, reduces to:

$$\int_{\Omega_S} \tilde{L}_\lambda(\mathbf{x}_S, \boldsymbol{\omega}) \cos \alpha_o \, d\omega \approx \tilde{L}_\lambda \cos \alpha_{os} \int_{\Omega_S} d\omega = \tilde{L}_\lambda \cos \alpha_{os} \, \Omega_S \qquad (24)$$

where $\Omega_S$ is the solid angle subtended by the source as seen from $\mathbf{x}_o$, and the approximate equality is consequence of the small size of the source (such that $\cos \alpha_o \approx \cos \alpha_{os}$ for all source points $\mathbf{x}_S$). Again under the small source assumption, this solid angle is equal to the transverse surface of the source projected on the direction joining $\mathbf{x}_o$ and $\mathbf{x}_S$, ($S_T$), divided by the square of the distance $R_o = \|\mathbf{x}_S - \mathbf{x}_o\|$, so that

$$\int_{\Omega_S} \tilde{L}_\lambda(\mathbf{x}_S, \boldsymbol{\omega}) \cos \alpha_o \, d\omega = \frac{\tilde{L}_\lambda \cos \alpha_{os} \, S_T}{R_o^2} \qquad (25)$$

Assuming a spherical source of radius $R_s$, we have $S = 4\pi R_s^2$ and $S_T = \pi R_s^2$, hence $S_T/S = 1/4$, and, from (23) and (25):

$$\int_{\Omega_S} \tilde{L}_\lambda(\mathbf{x}_S, \boldsymbol{\omega}) \cos \alpha_o \, d\omega = \frac{\Phi_S(\lambda)}{4\pi R_o^2} \cos \alpha_{os} \qquad (26)$$

which is, as expected, the irradiance produced by a pointlike isotropic source of flux $\Phi_S(\lambda)$ onto a point located a distance $R_o$ away, on a surface whose normal forms an angle $\alpha_{os}$ with the line joining the point and the source. Besides, since the source is factorizable, Eqs. (16) and (23):

$$B(\mathbf{x}, \boldsymbol{\omega}) = \frac{1}{\pi S} \, . \qquad (27)$$

Now, the linear operator series in Eqs. (11)-(12) can be easily evaluated. The zero-order term is given by:

$$G_0[\tilde{L}_\lambda(\mathbf{x}',\boldsymbol{\omega}');\mathbf{x},\boldsymbol{\omega}] = \tilde{L}_\lambda(\mathbf{x},\boldsymbol{\omega}) = g(\mathbf{x},\boldsymbol{\omega})\frac{\Phi_S(\lambda)}{\pi S}, \qquad (28)$$

where $g(\mathbf{x},\boldsymbol{\omega})$ is a geometrical factor equal to 1 if the initial point of the ray $(\mathbf{x},\boldsymbol{\omega})$ is located on the source, and 0 otherwise. The first-order term involves an integration over the solid angle $\Omega = \Omega_S$ subtended by the source as seen from the point $\mathbf{x}$ of the walls:

$$\begin{aligned} G_1[\tilde{L}_\lambda(\mathbf{x}',\boldsymbol{\omega}');\mathbf{x},\boldsymbol{\omega}] &= \int_\Omega \tilde{L}_\lambda(\mathbf{x}',\boldsymbol{\omega}')\, f_\lambda(\mathbf{x},\boldsymbol{\omega}',\boldsymbol{\omega})\cos\alpha\, d\omega' = \\ &= \frac{\Phi_S(\lambda)}{\pi S}\frac{\rho(\lambda)}{\pi}\int_\Omega g(\mathbf{x}',\boldsymbol{\omega}')\cos\alpha\, d\omega' \approx \frac{\Phi_S(\lambda)}{\pi S}\frac{\rho(\lambda)}{\pi}\int_\Omega g(\mathbf{x}',\boldsymbol{\omega}')\, d\omega' = \\ &= \frac{\Phi_S(\lambda)}{\pi S}\frac{\rho(\lambda)}{\pi}\Omega_S \approx \frac{\Phi_S(\lambda)}{\pi S}\frac{\rho(\lambda)}{\pi}\frac{S_T}{R^2} = \frac{1}{\pi}\frac{\Phi_S(\lambda)}{4\pi R^2}\rho(\lambda) \end{aligned} \qquad (29)$$

where successive use has been made of Eq. (28), conditions (iv) and (ii), the small source assumption such that, as seen from the walls, $\Omega_S \approx S_T/R^2$, and the spherical source assumption $S_T/S = 1/4$. Higher-order terms can immediately be computed taking into account that the integrations extend to the $\Omega = 2\pi$ hemisphere facing each wall point and that $\int_\Omega \cos\alpha\, d\omega = \pi$. Performing the integrations, we get, for the general $n$-th term:

$$G_n[\tilde{L}_\lambda(\mathbf{x}',\boldsymbol{\omega}');\mathbf{x},\boldsymbol{\omega}] = \frac{1}{\pi}\frac{\Phi_S(\lambda)}{4\pi R^2}\rho^n(\lambda). \qquad (30)$$

Summing up the infinite series in $\rho^n(\lambda)$, the radiance at any point $\mathbf{x}$ in the direction $\boldsymbol{\omega}$ is finally given by:

$$L_\lambda(\mathbf{x},\boldsymbol{\omega}) = \frac{\Phi_S(\lambda)}{\pi S}g(\mathbf{x},\boldsymbol{\omega}) + \frac{1}{\pi}\frac{\Phi_S(\lambda)}{4\pi R^2}\left[\frac{\rho(\lambda)}{1-\rho(\lambda)}\right]. \qquad (31)$$

The irradiance operator terms can be easily computed from Eq. (15) and Eqs.(28)-(30) as

$$F_0[\tilde{L}_\lambda(\mathbf{x}',\boldsymbol{\omega}');\mathbf{x}_o,\mathbf{n}_o] = \frac{\Phi_S(\lambda)}{\pi S}\int_{\Omega_S}\cos\alpha_o d\omega = \frac{\Phi_S(\lambda)}{\pi S}\cos\alpha_{os}\frac{S_T}{R_o^2} = \frac{\Phi_S(\lambda)}{4\pi R_o^2}\cos\alpha_{os}, \qquad (32)$$

for $n=0$, and

$$F_n[\tilde{L}_\lambda(\mathbf{x}',\boldsymbol{\omega}');\mathbf{x}_o,\mathbf{n}_o] = \frac{1}{\pi}\frac{\Phi_S(\lambda)}{4\pi R^2}\rho^n(\lambda)\int_\Omega \cos\alpha_o d\omega = \frac{\Phi_S(\lambda)}{4\pi R^2}\rho^n(\lambda),\qquad(33)$$

for $n\neq 0$. The irradiance, Eq. (14), is then given by:

$$E_\lambda(\mathbf{x}_o,\mathbf{n}_o) = \frac{\Phi_S(\lambda)}{4\pi R^2}\left[\left(\frac{R}{R_o}\right)^2\cos\alpha_{os} + \frac{\rho(\lambda)}{1-\rho(\lambda)}\right],\qquad(34)$$

and from this equation the analytic form of the effective inverse surface function, Eqs. (18) and (20), immediately follows:

$$T(\lambda;\mathbf{x}_o,\mathbf{n}_o) = \frac{1}{4\pi R^2}\left[\left(\frac{R}{R_o}\right)^2\cos\alpha_{os} + \frac{\rho(\lambda)}{1-\rho(\lambda)}\right].\qquad(35)$$

The first term of the *EISF* accounts for the direct contribution of the source to the observed irradiance: this term is in fact the inverse of the area of a sphere centered on the source and passing through the observer, $1/(4\pi R_o^2)$, multiplied by the cosine of the angle of incidence. The second term gives the contribution of the surfaces, taking into account the multiple ray reflections. The contribution of the first reflection in the walls is proportional to the reflectance $\rho(\lambda)$, whereas the total contribution of the reflected light depends on $\rho(\lambda)/[1-\rho(\lambda)]$, the sum of the infinite series of multiple reflections whose general term goes with $\rho^n(\lambda)$, with $n$=1 to infinity. The ratio of the single to the total reflected radiance is then of order $[1-\rho(\lambda)]$. Reflectances of indoor surfaces typically span a large range of values (~0.1-0.9). For an average reflectance $\rho(\lambda)$=0.5 the total contribution of the reflected light is expected to be twice the amount of light than underwent only the first reflection. In writing Eqs. (32)-(34) we have assumed that the light of the source is not obstructed by the observer in its way to the walls, and that there are no windows or other apertures in the homogeneous sphere surface.

Based on Eq. (34), we can define the Irradiance Amplification Factor (*IAF*) as the ratio of the irradiance on the observer's eye, gazing in arbitrary directions within the room, to the irradiance that would be obtained by looking directly at the source in the absence of walls. The *IAF* is given by:

$$IAF(\lambda) = \cos\alpha_{os} + \left(\frac{R_O}{R}\right)^2 \frac{\rho(\lambda)}{1-\rho(\lambda)} \qquad (36)$$

The minimum value of the *IAF* is obtained when the source is outside the hemispheric visual field of the observer (equivalent to setting $\cos\alpha_{os} = 0$), and the maximum is achieved when the observer looks directly at the source ($\cos\alpha_{os} = 1$)

To get some insight about the order of magnitude of the expected effects, we computed Eqs. (35) and (36) with a set of parameters close to the ones corresponding to the room in which the experimental measurements reported in section 4 were carried out. We used an equivalent radius $R = 2.5$ m, a distance source-observer $R_O = 1.3$ m and an equivalent spectral reflectance $\rho(\lambda)$ given by the weighted average of the measured spectral reflectances of the light and dark walls, ceiling, floor and table, with weighting factors proportional to the fractional room surface covered by each reflectance type. The results, for observers gazing directly at the source ($\alpha_{os} = 0°$), at intermediate directions ($\alpha_{os} = 52°$), and at the walls being the source behind the observer ($\alpha_{os} \geq 90°$), are shown in Fig. 10.

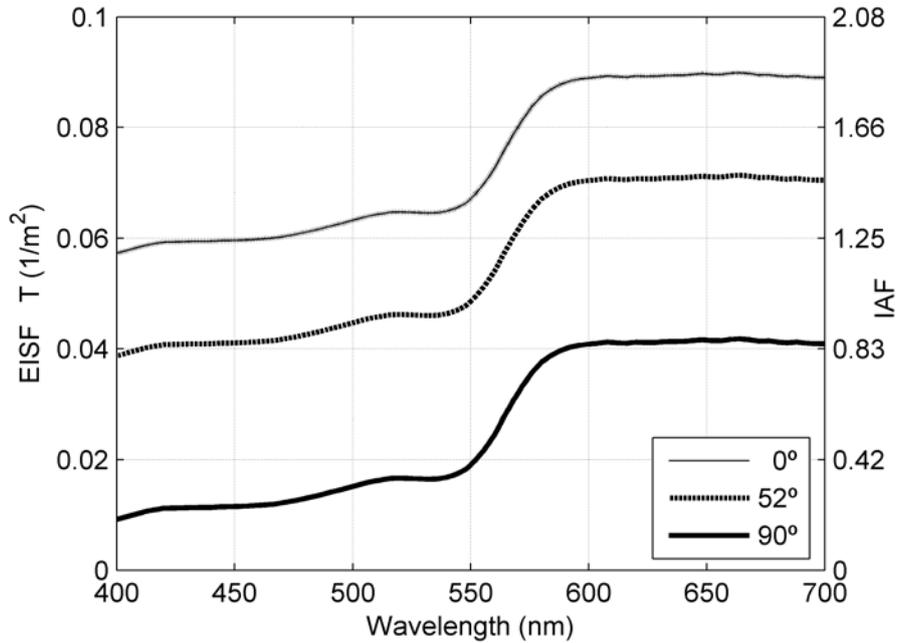

**Fig. 10.** Effective inverse surface functions [m$^{-2}$] (left axis) and irradiance amplification factors (dimensionless, right axis) for an observer in a simplified geometry (see text), with gazing directions $\alpha_{OS}$ = 0°, 52°, and $\geq$ 90°.

Although no accurate match is to be expected because the geometry and actual reflectances are necessarily different, the *EISF* behaviour in Fig. 10 is qualitatively similar to the one shown in Fig. 6. The *EISF* in each figure has essentially the same shape for different gaze directions, whereas its vertical position is shifted to a bigger or lesser extent depending on the angle of incidence of the direct light from the source on the cornea. The curve predicted in Fig. 10 for $\alpha_{OS} \geq 90°$ corresponds to the situation in which the observer looks towards the North wall, and it is remarkably close to that obtained in the actual room (Fig. 11), although the simplified analytical prediction tends to overestimate the measured EISF values by an average 25%. The curve for $\alpha_{OS}$ =52°, that corresponds to the experimental situation in which the observer looks towards the South wall, clearly underestimates the actual measurements (Fig. 6, lower right). Let us stress again that the main aim of the simplified model described in this section is to provide some analytical results that may capture the essential traits of the influence of the walls and other surrounding surfaces in the build up of the spectral irradiance at the entrance of the observer's eyes. It is intended mostly as a complementary aid for analyzing indoor exposures, and it is no substitute for the precise measurement of the EISF, or the numerical evaluation of the multiple reflection integrals described in sections 2 and 3.

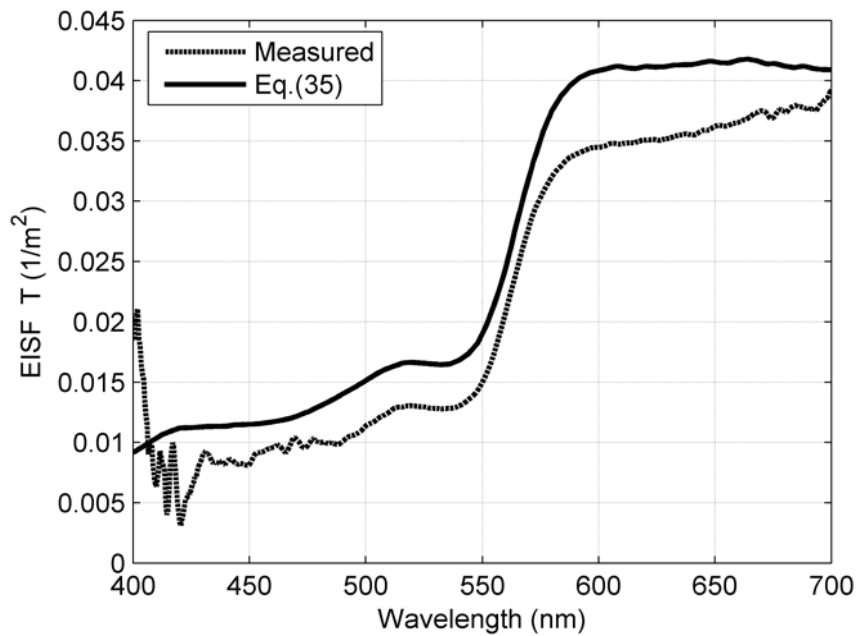

**Fig. 11.** Effective inverse surface functions [m$^{-2}$] for an observer gazing to the North wall in the experimental conditions described in the text. Dotted line: measured EISF (Fig. 6, upper left) Full line: EISF predicted using the simplified model of Eq. (35) (bottom curve of Fig 10) .

The *IAF* can reach values substantially higher than 1 for spectral regions of high reflectance. Indoor spaces, then, may reduce or enhance significantly the spectral content of the light reaching the observer's eyes in different spectral bands, in comparison with the direct light that would arrive from the source to the observer in the absence of walls.

## 6. Discusion

The radiance field is a useful tool for analyzing the light exposure of human observers in indoor spaces. If the sources are factorizable, the surrounding surfaces (wall, ceiling, floor, and objects within the room) act as an effective filter that selectively enhances or attenuates different regions of the source spectrum. This filter-like behaviour can be formalized through an effective inverse surface solid angle function, that once mutiplied by the spectral flux of the source straightforwardly provides the radiance field, and an effective inverse surface function that provides in an analogous way the spectral irradiance at the entrance of the eye of the observer for arbitrarily oriented gaze directions. These functions depend on the spatial and angular properties of the source radiance, on the room geometry and spectral reflectance, and on the observer position and gaze direction. Different sources, with different angular emission patterns, have to be described by different filter functions. However, these filter-like functions do not depend on the source spectrum, and hence they provide an easy, direct, and fast way of comparing the exposure conditions under light sources with different spectral compositions, everything else kept constant.

Taking into accout their construction technology and operating conditions, most indoor light sources could be considered factorizable. This condition has been checked in outdoor settings, providing a practicable method for retrieving the value of saturated spectral pixels in the context of hyperespectral analysis of urban nightscapes [37].

A straightforward analytical solution for the radiance field can be easily obtained for a particularly simple geometry, as described in Section 5. The detailed parameters are of course different from the ones existing at any realistic room; however, this model reproduces reasonably well the qualitative behavior of the measurements obtained in our experiments and may be helpful as a thinking aid. As a matter of fact, a typical indoor space essentially behaves as a very imperfect integrating sphere. This model provides some insights into the modulating role of the environment, and helps visualizing the selective attenuation or enhancement of different regions of the source spectrum. As commented above, the main value of the analytic solutions obtained for simplified ideal situations is that they provide useful physical insights on the phenomena under study. They are not expected to provide accurate numerical predictions, a task left for specific spectral rendering software. Numerical calculations with complex models are required for obtaining accurate radiance values in any actual experimental situation. Both approaches are complementary and do not intend to be substitute for each other.

Our results show that the wall reflectances can modify the spectral composition of the light reaching the human eyes in ways that may be relevant for the non-visual photic regulation of the human circadian system. In agreement with Bellia *et al* [21, 38], these results suggest that the indoor environment shall be taken into account when evaluating the relative merits of different light sources regarding its intended or unwanted non-visual effects.

Two observations are in place regarding metrics. On the one hand, the use of the CCT as a proxy for the lamp spectral power distribution is of course only approximate and shall be taken with due caution. It is the detailed spectral distribution, and not any single CCT metrics derived from it, what determines the inputs to the different eye photoreceptors: sources with the same CCT may produce significantly different photoreceptoral inputs. On the other hand, current phototransduction models are usually built in terms of the corneal irradiance, rather than of the corneal radiance: it is this last radiometric magnitude, however, the one that determines the spatial distribution of the irradiance across the retina and hence the amount of light captured by the photoreceptors at each retinal location. Any potential effects related with the balance between the central and peripheral retinal stimulations cannot be described in terms of the corneal irradiance alone.

The experimental results presented in this work correspond to a relatively simple indoor environment, an unsophisticated meeting room. Human beings are exposed daily to considerably more complex and varying surroundings. The particular situation here analyzed may approach one of the best-case scenarios, in terms of attenuating the short-wavelength regions of the spectrum by means of wall paints behaving as long-pass edge filters while still providing a pleasant visual atmosphere. In these conditions it can be achieved a reduction up to the 50% in the input to the ipRGC channel, at constant corneal illuminance (Fig. 8). Although this particular result cannot be directly extrapolated to other environments, we consider that it is indicative of the order of magnitude of the effects that can be expected due to the observer surroundings. A judicious choice of the wall paint reflectances, compatible with ensuring basic visual performance and fulfilling other design goals, may hence be useful for modulating the non-visual effects produced by light in domestic, workplace or clinical settings. It is increasingly frequent that people be exposed at night to the direct light of self-luminous displays of consumer electronic devices like tablets and smartphones. Since these devices are commonly held relatively close to the eye, their light arrives to the corneal plane without filtering, and it has ben shown to be able to produce measurable non-visual effects [23-24, 39-40]. A complete assessment of the human exposure to artificial light at night shall include when appropriate these direct sources, in addition to the static ambient lights.

As a final remark, the formal approach described in this paper could be useful to help bridging indoor and outdoor light pollution studies. The effects of light intrusion can be quantified by using the spectral radiance field produced by the light entering the room through its windows, acting as flat sources located at the walls, with a radiance distribution determined by outdoor streetlights, vehicle lights, and skyglow. And, conversely, the radiance field produced by indoor lamps can be evaluated at the window panes, quantifying that way the amount, directionality, and spectral distribution of the light that leaves the buildings and acts as a source of light pollution outdoors. These issues, that have not been developed here, deserve further research work.

**7. Conclusions**

The spectral radiance field is a basic tool for analyzing photic effects in indoor spaces. The general framework described in this paper enables addressing a wide range of problems related to the interactions between light souces, surrounding surfaces and human observers. For factorizable light sources, the modulating effects of the multiple wall reflections can be described in terms of a set of effective filter-like functions acting on the spectral flux distribution of the source.  As expected, the combination of selective absorption and multiple reflections at the surfaces surrounding the observer significantly enhances the radiance in some regions of the optical spectrum, while attenuating it in others. This gives rise to relevant changes in the spectral composition of the light reaching the observer eyes, in comparison with the one that would be produced by the light sources alone, in the absence of walls.

Since non-visual interactions are strongly wavelength dependent, a suitable design of the indoor spaces, particularly in what concerns the spectral reflectance of their main surfaces, offers an additional degree of freedom for controlling, within certain limits, the exposure of the observer to certain regions of the optical spectrum. In the reported experimental case, the use of wall paints acting as long pass filters allowed to decrease significantly the photic inputs to some of the five basic retinal photoreceptors as well as to reduce the melatonin suppression levels predicted by current phototransduction models. The expected changes produced by the environment include the existence of interesting sign reversals of the blue-yellow spectral opponency term.

Tayloring the spectral reflectance of the walls may provide additional degrees of freedom for the design of healthy indoor environments. Any such control, of course, must take also into account the lighting levels and the spectral conditions required for ensuring an adequate visual performance and achieving other desirable design goals. More research on the photic interactions of light is needed, however, before optimal detailed schemes of indoor lighting for human wellbeing can be standarized.

**Acknowledgments**

This work was developed within the framework of the Spanish Network for Light Pollution Studies (AYA2015-71542-REDT).

11. Brainard GC, Hanifin JP, Greeson JM, Byrne B, Glickman G, Gerner E, Rollag MD. Action spectrum for melatonin regulation in humans: evidence for a novel circadian photoreceptor. Journal of Neuroscience 2001; 21:6405–6412.
12. Thapan K, Arendt J, Skene DJ. An action spectrum for melatonin suppression: evidence for a novel non-rod, non-cone photoreceptor system in humans. Journal of Physiology 2001; 535:261–267.
13. Berson DM, Dunn FA, Takao M. Phototransduction by retinal ganglion cells that set the circadian clock. Science 2002; 295:1070–1073.
14. Dacey DM, Liao HW, Peterson BB, Robinson FR, Smith VC, Pokorny J, Yau KW, Gamlin PD. Melanopsin-expressing ganglion cells in primate retina signal colour and irradiance and project to the LGN. Nature 2005; 433:749–754.
15. Brainard GC, Sliney D, Hanifin JP, Glickman G, Byrne B, Greeson JM, Jasser S, Gerner E, Rollag MD. Sensitivity of the human circadian system to short-wavelength (420-nm) light. Journal of Biological Rhythms 2008; 23:379–386.
16. Zukauskas A, Vaicekauskas R, Vitta P. Optimization of solid-state lamps for photobiologically friendly mesopic lighting. Applied Optics 2011; 51:8423–8432.
17. Falchi F, Cinzano P, Elvidge CD, Keith DN, Haim A. Limiting the impact of light pollution on human health, environment and stellar visibility. Journal of Environmental Management 2011; 92:2714–2722.
18. Czeisler CA. Perspective: Casting light on sleep deficiency. Nature 2013; 497:S13.
19. Aubé M, Roby J, Kocifaj M. Evaluating Potential Spectral Impacts of Various Artificial Lights on Melatonin Suppression, Photosynthesis, and Star Visibility. PLoS ONE 2013; 8:e67798.
20. Dijk DJ. Why do we sleep so late?. Journal of Sleep Research 2013; 22:605–606.
21. Bellia L, Pedace A, Barbato G. Indoor artificial lighting: Prediction of the circadian impact of different spectral power distributions. Lighting Research and Technology 2014: 46:650–660.
22. Bonmati-Carrion MA, Arguelles-Prieto R, Martinez-Madrid MJ, Reiter R, Hardeland R, Rol MA, Madrid JA. Protecting the melatonin rhythm through